\documentclass[amsmath,11pt]{article}

\usepackage{amsfonts,amsmath,amssymb,graphics,epsfig}

\date{\empty}

\begin{document}

\title{\bf Relativistic magnetised perturbations: magnetic pressure vs magnetic tension}

\author{Dimitra Tseneklidou${}^1$, Christos G. Tsagas${}^1$ and John D. Barrow${}^2$\\ {\small ${}^1$Section of Astrophysics, Astronomy and Mechanics, Department of Physics}\\ {\small Aristotle University of Thessaloniki, Thessaloniki 54124, Greece}\\ {\small ${}^2$DAMTP, Centre for Mathematical Sciences, University of Cambridge}\\ {\small Wilberforce Road, Cambridge CB3 0WA, United Kingdom}}

\maketitle

\begin{abstract}
We study the linear evolution of magnetised cosmological perturbations in the post-recombination epoch. Using full general relativity and adopting the ideal magnetohydrodynamic approximation, we refine and extend the previous treatments. More specifically, this is the first relativistic study that accounts for the effects of the magnetic tension, in addition to those of the field's pressure. Our solutions show that on sufficiently large scales, larger than the (purely magnetic) Jeans length, the perturbations evolve essentially unaffected by the magnetic presence. The magnetic pressure dominates on small scales, where it forces the perturbations to oscillate and decay. Close to the Jeans length, however, the field's tension takes over and leads to a weak growth of the inhomogeneities. These solutions clearly demonstrate the opposing action of the aforementioned two magnetic agents, namely of the field's pressure and tension, on the linear evolution of cosmological density perturbations.
\end{abstract}

\section{Introduction}\label{sI}
Large-scale magnetic ($B$) fields appear to be quite common in our universe (e.g.~see~\cite{CT}-\cite{V}), with a verified presence in stars, galaxies, galaxy clusters, high-redshift protogalaxies and possibly even in intergalactic voids. Nevertheless, the origin, the evolution and the role of these large-scale cosmic magnetic fields remain essentially unknown, although their widespread presence might suggest that they are of primordial nature. The case for cosmological magnetic fields got stronger when recent reports claimed the existence of coherent magnetic fields in the low density intergalactic space (where no dynamo amplification is likely to operate) with strengths around $10^{-15}~G$~\cite{AK}-\cite{CBF}. Additional support comes from the fact that galaxies (like our Milky Way), galaxy clusters and remote protogalaxies have $B$-fields of similar ($\mu G$-order) strengths, which could be a sign of a common origin for all these fields.

It has long been known that large-scale cosmological magnetic fields, if present, would have affected the evolution of (baryonic) density perturbations, during both the linear and the non-linear regime of structure formation. More specifically, the presence of the $B$-field is believed to slow down the standard growth-rate of linear density gradients by an amount proportional to the square of the Alv\'en speed. Nevertheless, the available Newtonian and relativistic studies (see~\cite{RR}-\cite{TS} and~\cite{TB1}-\cite{BMT} respectively) account only for the contribution of the magnetic pressure, namely of the field's positive pressure. The effects of the magnetic tension, that is of the negative pressure exerted along the field lines themselves, have never been accounted for. The only exception has been a recent Newtonian study, where the role of the magnetic tension on the linear evolution of density inhomogeneities in the post-recombination universe was investigated~\cite{VT}. That work indicated that the aforementioned two magnetic agents may have opposing action, but the results were not conclusive.

Here, we provide the first (to the best of our knowledge) fully relativistic study of magnetised density perturbations that incorporates the effects of the field's tension, in addition to those of its (positive) pressure. Our starting point is a perturbed, nearly flat, Friedmann-Robertson-Walker (FRW) universe permeated by a weak large-scale magnetic field. The latter could be primordial in origin, or a later addition to the phenomenology of our universe (e.g.~see~\cite{KKT,Wetal} for recent reviews). Confining to the post-recombination epoch, where structure formation starts in earnest, we set the matter pressure to zero and focus on the role and the implications of the $B$-field. The latter affects the linear evolution of density inhomogeneities through the Lorentz force, which splits into a pressure and a tension part. Not surprisingly, since we are dealing with dust, the magnetic pressure becomes the sole source of support against the gravitational pull of the matter. The linear contribution of the magnetic tension, on the other hand, is two-fold. There are pure-tension stresses, similar but not identical to those identified in the Newtonian study of~\cite{VT}, and a purely relativistic magneto-curvature stress triggered by the non-Euclidean geometry of the host space. Both of these tension stresses reflect the elasticity of the magnetic forcelines and their generic tendency to react against any agent (physical or geometrical) that distorts them from equilibrium~\cite{P}-\cite{T}.

We analyse the role of the $B$-field in a step-by-step approach, accounting for the effects of the magnetic pressure first, before gradually incorporating those of the field's tension. In the first instance, our results recover those of the earlier studies. We confirm that, when dealing with dust, there is a purely magnetic Jeans length below which density perturbations cannot grow. Instead, the density gradients oscillate with an amplitude that decays to zero. Well outside the aforementioned Jeans scale, on the other hand, the perturbations grow essentially unimpeded by the field's presence. Incorporating the effects of the magnetic tension does not seem to affect the large-scale evolution of the density gradients, since they continue to grow as if there was no $B$-field present. On wavelengths near and below the Jeans threshold, however, the standard picture changes when the tension stresses are accounted for. Although the density perturbations still oscillate with decreasing amplitude on scales well inside the Jeans length, they now decay to a finite (rather than to zero) amplitude. Moreover, close to the Jeans length, where the magnetic pressure balances out the gravitational pull of the matter and the field's tension becomes the main player, the perturbations experience a slow (logarithmic) growth. Despite the weakness of the effect, this result clearly demonstrates the opposing action of the aforementioned two magnetic agents and reveals the, as yet unknown, role of the magnetic tension.

In our final step we also incorporate the magneto-curvature (tension) stresses into the linear equations. However, our assumption of a spatially flat FRW background means that the associated effects are (by default) too weak to make a ``visible'' difference. The magneto-curvature stresses identified here could (in principle) play the dominant role during a curvature-dominated era, which could occur in the very early or in the very late evolution of the universe. In such a case the type of spatial curvature (i.e.~whether it is positive or negative) is of paramount importance, because it determines the nature of the magneto-geometrical effect.  Nevertheless, to look into the possible implications of such stresses, one needs to consider cosmological backgrounds with nonzero spatial curvature, which goes beyond the scope of the present work.

\section{Relativistic magnetohydrodynamics}\label{sRMHD}
We will study the role of cosmological magnetic fields on density perturbations by applying the 1+3 covariant formalism to relativistic cosmic magnetohydrodynamics (MHD), an approach that has been proven a powerful tool in the past (e.g.~see~\cite{TB1}-\cite{BMT}).

\subsection{The 1+3 spacetime splitting}\label{ss1+3SS}
Introducing a family of timelike observers allows for the 1+3 threading of the spacetime into time and 3-dimensional space. The temporal direction is defined by the observers' 4-velocity field ($u_a$ -- normalised so that $u^{a}u_{a}=-1$), while their rest-space is associated with the projector $h_{ab}=g_{ab}+ u_{a}u_{b}$ (where $g_{ab}$ is the spacetime metric). The latter is a symmetric tensor that projects orthogonally to the 4-velocity vector (i.e.~$h_{ab}u^b=0$) and acts as the metric of the 3-space when there is no rotation. Using both $u_a$ and $h_{ab}$, we define the temporal and spatial derivatives of a general tensor field $S_{ab...}{}^{cd...}$ as
\begin{equation}
\dot{S}_{ab\cdots}{}^{cd\cdots}= u^{e}\nabla_{e}S_{ab\cdots}{}^{cd\cdots} \hspace{10mm} {\rm and} \hspace{10mm} {\rm D}_{e}S_{ab\cdots}{}^{cd\cdots}= h_e{}^{s}h_a{}^{f}h_b{}^{p}h_q{}^{c}h_r{}^{d}\cdots \nabla_{s}S_{fp\cdots}{}^{qr\cdots}\,,  \label{dD}
\end{equation}
respectively (with $\nabla_a$ representing the 4-dimensional covariant derivative operator).

\subsection{Matter fields and kinematics}\label{ssMFKs}
The 1+3 formalism utilises a systematic decomposition of all physical variables and operators into their irreducible temporal and spatial components. For instance, relative to the observers introduced earlier, which in our case are always comoving with the matter, the energy-momentum tensor of a general imperfect fluid splits as
\begin{equation}\label{Tab}
T_{ab}= \rho u_au_b+ ph_{ab}+ 2q_{(a}u_{b)}+ \pi_{ab}\,,
\end{equation}
with $\rho=T_{ab}u^au^b$, $p=T_{ab}h^{ab}/3$, $q_a=-h_a{}^bT_{bc}u^c$ and $\pi_{ab}=h_{\langle a}{}^ch_{b\rangle}{}^dT_{cd}$ representing the energy density, the isotropic pressure, the energy flux and the viscosity of the matter respectively.\footnote{Square brackets indicate antisymmetrisation, round ones symmetrisation and angled brackets denote the symmetric and traceless part of second rank spacelike tensors. For instance, $\pi_{ab}=h_{(a}{}^ch_{b)}{}^dT_{cd}-ph_{ab}$. Also, we use geometrised and Heaviside-Lorentz units throughout this manuscript.} Note that the quantities on the right-hand side of the above correspond to the total matter, which may also include electromagnetic fields.

In an analogous way, the covariant derivative of the observers' 4-velocity decomposes into the irreducible kinematic variables according to
\begin{equation}\label{velocity gradient}
\nabla_{b}u_{a}= \frac{1}{3}\,\Theta h_{ab}+ \sigma_{ab}+ \omega_{ab}- A_{a}u_{b}\,,
\end{equation}
where $\Theta=\nabla^a u_a={\rm D}^au_a$ is the volume expansion/contraction scalar, $\sigma_{ab}={\rm D}_{\langle b}u_{a\rangle}$ is the shear tensor, $\omega_{ab}={\rm D}_{[b}u_{a]}$ is the vorticity tensor and $A_a=\dot{u}_a$ is the 4-acceleration vector. By construction, $\sigma_{ab}u^a=0= \omega_{ab}u^a=A_{a}u^a$, which ensures that all three of them are spacelike. The volume scalar monitors the mean separation between neighbouring observers and it is also used to introduce a characteristic length-scale along their worldlines. This is the cosmological scale-factor ($a$) defined by means of $\dot{a}/a=\Theta/3$. The shear and the vorticity describe kinematic anisotropies and rotation respectively, while the 4-acceleration implies the presence of non-gravitational forces. Note that the antisymmetry of the vorticity tensor means that it can be replaced by the vector $\omega_a=\varepsilon_{abc}\omega^{bc}/2$, where $\varepsilon_{abc}$ is the Levi-Civita tensor of the 3-space.

The evolution of the kinematic variables defined above follows after applying the Ricci identities to the 4-velocity vector, namely from the expression $2\nabla_{[a}\nabla_{b]}u_c=R_{abcd}u^d$, where $R_{abcd}$ is the Riemann tensor of the spacetime. The Ricci identities decompose into a set of three timelike formulae, monitoring the evolution of the $u_a$-field, and into an equal number of spacelike relations that act as constraints. Referring the reader to~\cite{TCM} for further discussion and details, we will only provide the propagation equation of the volume scalar, namely
\begin{equation}\label{Raychaudhuri}
\dot{\Theta}=-\frac{1}{3}\,\Theta^{2}- \frac{1}{2}\,(\rho+3p)- 2\left(\sigma^2-\omega^2\right)+ {\rm D}^{a}A_{a}+ A^{a}A_{a}+ \Lambda\,,
\end{equation}
with $\Lambda$ representing the cosmological constant, $\sigma^2= \sigma_{ab}\sigma^{ab}/2$ and $\omega^2= \omega_{ab}\omega^{ab}/2$ by definition. The above expression, which is commonly known as the Raychaudhuri equation, applies to a general spacetime filled with an imperfect fluid of arbitrary electrical conductivity.

\subsection{Magnetohydrodynamics and conservation laws}\label{ssMHDCLs}
The post-inflationary universe is treated as a very good electrical conductor, at least on subhorizon scales where causal microphysical processes readily apply (see also footnote~2 below). In such an environment, the ideal-MHD limit is believed to provide an excellent physical approximation. Mathematically speaking this means setting $\varsigma\rightarrow\infty$, where $\varsigma$ is the electrical conductivity of the cosmic medium. In a frame comoving with the matter, the covariant form of Ohm's law reads $\mathcal{J}_a=\varsigma E_a$, with $\mathcal{J}_a$ and $E_a$ representing the spatial currents and the electric field respectively (e.g.~see~\cite{J}). Then, in the presence of finite 3-currents, $E_a\rightarrow0$ as $\varsigma\rightarrow\infty$. All these ensure that at the ideal-MHD limit, Maxwell's equations reduce to one propagation formula
\begin{equation}\label{Faraday}
\dot{B}_{\langle a\rangle}= -{2\over3}\,\Theta B_a+ \left(\sigma_{ab}+\epsilon_{abc}\omega^c\right)B^b
\end{equation}
and  three constraints
\begin{equation}\label{Gauss,Coulomb,Ampere}
{\rm curl}B_a= \mathcal{J}_a- \varepsilon_{abc}A^bB^c\,, \hspace{10mm} 2\omega_aB^a= \mu \hspace{10mm} {\rm and} \hspace{10mm} {\rm D}^aB_a= 0\,,
\end{equation}
where $\dot{B}_{\langle a\rangle}=h_a{}^b\dot{B}_b$, ${\rm curl}B_a=\varepsilon_{abc}{\rm D}^bB^c$ and $\mu$ is the electric charge density of the matter~\cite{TB1}-\cite{BMT}. The former of the above, namely Eq.~(\ref{Faraday}), guarantees that the magnetic forcelines always connect the same particles at all times~\cite{E1}. In other words, at the ideal MHD approximation, the $B$-field is frozen into the highly conductive matter.

In the absence of electric fields, the total energy-momentum tensor of the magnetised matter, assuming that the latter is a perfect fluid of arbitrarily high electrical conductivity, reads
\begin{equation}\label{Ttot}
T_{ab}= \left(\rho+\frac{1}{2}\,B^2\right)u_au_b+ \left(p+\frac{1}{6}\,B^2\right)h_{ab}+ \Pi_{ab}\,,
\end{equation}
where $B^2=B_aB^a$ and $\Pi_{ab}=\Pi_{\langle ab\rangle}= (B^2/3)h_{ab}-B_aB_b$. The former provides a measure of the magnetic energy density and isotropic pressure, while the latter defines the anisotropic pressure of the $B$-field~\cite{TB1}-\cite{BMT}. Following (\ref{Tab}) and (\ref{Ttot}), we deduce that our magnetised medium behaves as an imperfect fluid with effective energy density $\rho+B^2/2$, effective isotropic pressure $p+B^2/6$ and effective viscosity $\Pi_{ab}$. The latter is a symmetric and trace-free spacelike tensor, which unveils the generically anisotropic nature of the $B$-field. Note that $\Pi_{ab}$ has positive eigenvalues orthogonal to the magnetic forcelines and negative parallel to them. More specifically, it is straightforward to show that $\Pi_{ab}n^b=(B^2/3)n_a$ and that $\Pi_{ab}k^b=-(2B^2/3)k_a$, where $n_a$ and $k_a$ are the unit vectors normal and along $B_a$ respectively. The positive eigenvalues are associated with the ordinary magnetic pressure and reflect the tendency of the field lines to fend off. The negative eigenvalue, on the other hand, manifests the tension properties of the magnetic forcelines, their elasticity and their intrinsic ``preference'' to remain as straight as possible~(e.g.~see~\cite{P,M}).

The conservation laws for the energy and the momentum densities of a highly conductive magnetised fluid follow from the (twice contracted) Bianchi identities and from Maxwell's equations. In particular, assuming a perfect medium, the timelike and the spacelike parts of the aforementioned Bianchi identities lead to the energy density
\begin{equation}\label{rho}
\dot{\rho}= -\Theta(\rho+p)
\end{equation}
and to the momentum-density
\begin{equation}\label{A}
(\rho+p)A_a= -{\rm D}_ap- \varepsilon_{abc}B^b\mathcal{J}^c\,,
\end{equation}
conservation laws, namely to the continuity equation and to the Navier-Stokes equation respectively.\footnote{Following (\ref{A}), the magnetic effects on the fluid propagate via the Lorentz force and require the presence of coherent electric currents. These are generated after inflation, which means that their size cannot exceed that of the causal horizon. Therefore, the magnetic effects discussed in this work apply primarily to subhorizon scales.} At the same time, the induction equation (see relation (\ref{Faraday}) above) leads to the conservation law of the magnetic energy density, namely to~\cite{TB1}-\cite{BMT}
\begin{equation}\label{dotB2}
\left(B^2\right)^{\cdot}= -{4\over3}\,\Theta B^2- 2\sigma_{ab}\Pi^{ab}\,.
\end{equation}
Expressions (\ref{rho}) and (\ref{dotB2}) reveal that, at the ideal-MHD limit, the energy density of the magnetised matter and that of the $B$-field itself are separately conserved.

\section{Magnetised density inhomogeneities}\label{ssMDIs}
Inhomogeneities in the density distribution of the matter are affected by pressure gradients. As mentioned above, the magnetic field is an additional source of pressure, both positive and negative. In what follows we will study the implications of these two different types of pressure for the linear evolution of magnetised density perturbations in the post-recombination universe.

\subsection{The key variables}\label{ssKVs}
Following the earlier relativistic treatments of~\cite{TB1,TM} (see also~\cite{BMT} for a review), we monitor inhomogeneities in the density distribution of matter by means of the dimensionless gradient
\begin{equation}\label{Drel}
\Delta_a= \frac{a}{\rho}\,{\rm D}_a\rho\,.
\end{equation}
The above variable, which depicts spatial variations in the matter density as measured by a pair of neighbouring observers, is supplemented by the auxiliary quantities
\begin{equation}\label{ZBrel}
\mathcal{Z}_a= a{\rm D}_a\Theta \hspace{10mm} {\rm and} \hspace{10mm} \mathcal{B}_a= \frac{a}{B^2}\,{\rm D}_aB^2\,.
\end{equation}
These, in turn, monitor local inhomogeneities in the volume expansion and in the magnetic energy density respectively. Note that all of the above vanish identically in an FRW background (see \S~\ref{ssBM} next) and for this reason they are gauge-invariant linear perturbations~\cite{SW}.

\subsection{The background model}\label{ssBM}
Our aim is to study the magnetic implications for the evolution of density perturbations in a perturbed almost-FRW universe. We therefore select as our background model a spatially flat Friedmann model with zero cosmological constant. Also, to enhance the linear magnetic effects, we will allow for the presence of completely random and sufficiently weak background magnetic field. The randomness implies that $\langle B_a\rangle=0$, which preserves the isotropy of the FRW host, while $\langle B^2\rangle\neq0$. The weakness ensures that, although the $B$-field contributes to the background energy density, its input is small (i.e.~$\langle B^2\rangle\ll\rho$), leaving the standard FRW dynamics unaffected.

The symmetries of the Friedmannian spacetimes imply that the only surviving background variables are time-dependent scalars. All the rest vanish identically and they will be therefore treated as first-order (gauge-invariant) perturbations. These include, among others, the inhomogeneity variables introduced in \S~\ref{ssKVs} earlier. Then, using overbars to denote the zero-order quantities, while setting $\bar{\Theta}=3H$ (where $H=\dot{a}/a$ is the unperturbed Hubble parameter) and $\bar{B}^2=\langle B^2\rangle$, with $\bar{B}^2=\bar{B}^2(t)$, the background evolution is monitored by the set
\begin{equation}\label{bgrFr}
3H^2= \bar{\rho}\,, \hspace{10mm} \dot{H}= -H^2- \frac{1}{6}\left(\bar{\rho}+3\bar{p}\right)\,, \hspace{10mm} \dot{\bar{\rho}}= -3H\left(\bar{\rho}+\bar{p}\right)
\end{equation}
and
\begin{equation}\label{bgrB2}
(\bar{B}^2)^\cdot= -4H\bar{B}^2\,.
\end{equation}
Note that we have ignored the magnetic contribution to the zero-order Friedmann and Raychaudhuri equations (see expressions (\ref{bgrFr}a) and (\ref{bgrFr}b) above), given that $\bar{B}^2\ll\bar{\rho}$ in the background. We also remind the reader that, at the ideal MHD limit, the energy density of the matter and that of the $B$-field are separately conserved (see Eqs.~(\ref{bgrFr}c) and (\ref{bgrB2})). The latter relation also unveils the radiation-like evolution of the zero-order magnetic field, namely that $\bar{B}^2\propto a^{-4}$, which also guarantees magnetic-flux conservation.

\subsection{Linear evolution of the inhomogeneities}\label{ssLEIs}
The nonlinear formulae describing the general evolution of magnetised density inhomogeneties can be found in~\cite{TB1}-\cite{BMT}, where we refer the reader for further discussion and technical details. Here, we will linearise these relations around a spatially flat FRW background (with zero cosmological constant) permeated by a sufficiently random and weak magnetic field (see \S~\ref{ssBM} before). In doing so, we will treat the magnetic energy-density and pressure gradients as first-order perturbations, which makes the perturbed $B$-field (and its spatial gradients) half-order perturbations.\footnote{The magnetic contribution to the linear equations comes always through terms of order $B^2$ (see expressions (\ref{linD'rel})-(\ref{linB'rel})), which ensures the perturbative consistency of the adopted linearisation scheme.} On these grounds, the linear evolution of the inhomogeneities is monitored by the propagation formulae~\cite{BMT}
\begin{equation}\label{linD'rel}
\dot{\Delta}_a= 3wH\Delta_a- (1+w)\mathcal{Z}_a+ \frac{3aH}{\bar{\rho}}\,\varepsilon_{abc}B^b{\rm curl}B^c+ 2aH(1+w)c_{\rm a}^2A_a\,,
\end{equation}
\begin{equation}\label{linZ'rel}
\dot{\mathcal{Z}}_a= -2H\mathcal{Z}_a- \frac{1}{2}\,\bar{\rho}\Delta_a- \frac{1}{2}\,\bar{\rho}(1+w)c_{\rm a}^2\mathcal{B}_a+ \frac{3a}{2}\,\varepsilon_{abc}B^b{\rm curl}B^c+ a{\rm D}_a{\rm D}^bA_b
\end{equation}
and
\begin{equation}\label{linB'rel}
\dot{\mathcal{B}}_a= \frac{4}{3(1+w)}\,\dot{\Delta}_a- \frac{4wH}{1+w}\,\Delta_a- \frac{4aH}{\bar{\rho}(1+w)}\,\varepsilon_{abc}B^b{\rm curl}B^c- 4aHA_a\,.
\end{equation}
According to (\ref{linD'rel}) and \ref{linZ'rel})), the magnetic field also sources inhomogeneities, both in the density of the matter and in the volume expansion of the universe. In the above $w=\bar{p}/\bar{\rho}$ is the background barotropic index of the matter and $c_{\rm a}^2=\bar{B}^2/\bar{\rho}(1+w)$ defines the zero-order Alfv\'{e}n speed. By construction, the latter satisfies the constraint $c_{\rm a}^2\ll1$ due to the overall weakness of the $B$-field. Finally, to linear order, the 4-acceleration vector seen on the right-hand side of the above is given by the momentum conservation law (see Eq.~(\ref{A}) in \S~\ref{ssMHDCLs}), which now reads
\begin{equation}
\rho(1+w)A_a= -{\rm D}_ap- \varepsilon_{abc}B^b{\rm curl}B^c= {\rm D}_ap- {1\over2}\,{\rm D}_aB^2+ B^b{\rm D}_bB_a\,,  \label{MHDAa}
\end{equation}
since $J_a={\rm curl}B_a=\varepsilon_{abc}{\rm D}^bB^c$ to linear order (see Eq.~(\ref{Gauss,Coulomb,Ampere}a)). Note that in the second equality of the above the Lorentz force splits into its pressure and tension stresses, given by ${\rm D}_aB^2/2$ and $B^b{\rm D}_bB_a$ respectively. In what follows, we will investigate the implications of these two magnetic agents for the evolution of linear perturbations in the density distribution of the matter.

\subsection{Types of inhomogeneities}\label{ssTIs}
The variables defined in \S~\ref{ssKVs}, namely $\Delta_a$, $\mathcal{Z}_a$ and $\mathcal{B}_a$, contain collective information about three types of inhomogeneities: scalar, vector and tensor. The former monitors overdensities or underdensities in the matter distribution, which we usually refer to as density perturbations. Vector inhomogeneities, on the other hand, describe rotational (vortex-like) distortions in the matter. Finally, tensor inhomogeneities describe changes in the shape of the density profile under constant volume. We may decode all this information by taking the comoving spatial gradient of $\Delta_a$ and then implementing the irreducible decomposition (e.g.~see~\cite{TCM})
\begin{equation}\label{Delab}
\Delta_{ab}= a{\rm D}_b\Delta_a= {1\over3}\,\Delta h_{ab}+ \Sigma_{ab}+ \varepsilon_{abc}W^c\,.
\end{equation}
Here $\Delta=a{\rm D}^a\Delta_a$ is the scalar describing overdensities/underdensities in the matter, $W_a=-a{\rm curl} \Delta_a/2$ is the vector monitoring density vortices and $\Sigma_{ab}=a{\rm D}_{\langle b}\Delta_{a\rangle}$ is the symmetric and trace-free tensor following changes in the shape of the density profile. Clearly, similar decompositions also apply to the expansion and the magnetic energy-density gradients~\cite{TM,BMT}.

The anisotropic nature of the $B$-field ensures its interaction with all of the aforementioned three types of inhomogeneities. Here, we will focus on the linear evolution of scalar density perturbations after recombination. This restriction means that we may set the matter pressure and the associated barotropic index to zero.

\section{Magnetised density perturbations}\label{sMDPs}
When dealing with dust, the magnetic field becomes the sole source of pressure support. However, this does not a priori guarantee that the growth of density perturbations will slow down in the magnetic presence, since the $B$-field is a source of negative pressure (tension) as well.

\subsection{Linear evolution of density perturbations}\label{ssLEDPs}
Scalar perturbations in the matter density, in the volume expansion of the universe and in the magnetic energy density are monitored by
\begin{equation}\label{scalarsD,Z,B}
\Delta= a{\rm D}^a\Delta_a\,, \hspace{10mm}    \mathcal{Z}= a{\rm D}^a\mathcal{Z}_a\,,  \hspace{10mm} {\rm and} \hspace{10mm}    \mathcal{B}= a{\rm D}^a\mathcal{B}_a\,,
\end{equation}
respectively (see \S~\ref{ssTIs} above). Then, setting $w=0$ and $c_{\rm a}^2=\bar{B}^2/\bar{\rho}\ll1$, the comoving 3-divergences of Eqs.~(\ref{linD'rel}), (\ref{linZ'rel}) and (\ref{linB'rel}) lead to the linear propagation formulae
\begin{equation}\label{scalarD'rel}
\dot{\Delta}= -\mathcal{Z}+ \frac{3}{2}\,Hc_{\rm a}^2\mathcal{B}- Hc_{\rm a}^2\mathcal{K}- \frac{6a^2H}{\bar{\rho}} \left(\sigma_B^2-\omega_B^2\right)\,,
\end{equation}
\begin{eqnarray}\label{scalarZ'rel}
\nonumber \dot{\mathcal{Z}}&=& -2H\mathcal{Z}- \frac{1}{2}\,\bar{\rho}\Delta+ \frac{1}{4}\,\bar{\rho}c_{\rm a}^2\mathcal{B}- \frac{1}{2}\,c_{\rm a}^2{\rm D}^2\mathcal{B}- \frac{1}{2}\,\bar{\rho}c_{\rm a}^2\mathcal{K}- 3a^2\left(\sigma_B^2-\omega_B^2\right) \\&& +\frac{2a^2}{\bar{\rho}}\,{\rm D}^2\left(\sigma_B^2-\omega_B^2\right)
\end{eqnarray}
and
\begin{equation}\label{scalarB'rel}
\dot{\mathcal{B}}= \frac{4}{3}\,\dot{\Delta}\,,
\end{equation}
respectively.\footnote{In deriving Eq.~(\ref{scalarZ'rel}) we have also used the linear auxiliary relation
\begin{equation}\label{scalar A}
a^2A= -\frac{1}{2}\,c_{\rm a}^2\mathcal{B}+ \frac{1}{3}\,c_{\rm a}^2\mathcal{K}+ \frac{2a^2}{\bar{\rho}}\left(\sigma_B^2-\omega_B^2\right)\,,
\end{equation}
where $A={\rm D}_aA^a$ is the 3-divergence of the 4-acceleration. Note that the last two terms on the right-hand side of the above, together with the last two terms of (\ref{scalarD'rel}) and the last three terms of (\ref{scalarZ'rel}), represent tension stresses, the effects of which were not included in the relativistic solutions of~\cite{TB1}-\cite{BMT}. The Newtonian analogues of the magnetic shear and vorticity, on the other hand, were accounted for in~\cite{VT} (see Eqs.~(23) and (27) there).} Note that the scalars $\sigma_B^2={\rm D}_{\langle b}B_{a\rangle}{\rm D}^{\langle b}B^{a\rangle}/2$ and $\omega_B^2={\rm D}_{[b}B_{a]}{\rm D}^{[b}B^{a]}/2$ are respectively related to shape and rotational distortions in a field-line congruence, which makes the tensors $\sigma_{ab}^B= {\rm D}_{\langle b}B_{a\rangle}$ and $\omega_{ab}^B= {\rm D}_{[b}B_{a]}$ the magnetic analogues of the kinematic shear and vorticity (see \S~\ref{ssMFKs}). Also, $\mathcal{K}=a^2\mathcal{R}$ by definition, with $\mathcal{R}$ representing the perturbed 3-Ricci scalar, which means that the fifth term on the right-hand side of Eq.~(\ref{scalarZ'rel}) carries the combined effects of magnetism and spatial curvature. Note that this particular stress reflects the vector nature of the $B$-field and derives from a purely geometrical coupling between magnetism and spacetime curvature~\cite{TM,T}. The latter comes into play via the Ricci identities and adds to the standard interaction between matter and geometry that the Einstein field equations introduce. Following~\cite{BMT}, the rescaled 3-Ricci scalar evolves as
\begin{equation}
\dot{\mathcal{K}}= -{4\over3}\,Hc_{\rm a}^2\mathcal{K}+ 2Hc_{\rm a}^2\mathcal{B}\,,  \label{lcK}
\end{equation}
to linear order. Finally, we should point out that the second term on the right-hand side of Eq.~(\ref{scalarD'rel}) and the third and fourth terms on the right-hand of (\ref{scalarZ'rel}) are due to the (positive) magnetic pressure. On the other hand, the last two terms of (\ref{scalarD'rel}) and the last three terms of (\ref{scalarZ'rel}) carry the effects of the field's tension.

\subsection{The wave-like equation}\label{ssW-LE}
Taking the time derivative of (\ref{scalarD'rel}), using the rest of the propagation formulae and keeping up to linear order terms, we obtain the following wave-like equation for the density perturbations
\begin{equation}\label{D''}
\ddot{\Delta}= -2H\dot{\Delta}+ \frac{1}{2}\,\bar{\rho}\Delta+ \frac{2}{3}\,c_{\rm a}^{2}{\rm D}^{2}\Delta+ \frac{2}{3}\,c_{\rm a}^{2}\rho\mathcal{K}+ 4a^{2}\left(\sigma_B^2-\omega_B^2\right)- \frac{2a^2}{\bar{\rho}}\,{\rm D}^2\left(\sigma_B^2-\omega_B^2\right)\,.
\end{equation}
with additional terms due to the universal expansion, the presence of matter (including the $B$-field) and spacetime curvature. In deriving the above we have also used the linear propagation formulae $(\sigma_B^2)^{\cdot}=-6H\sigma_B^2$ and $(\omega_B^2)^{\cdot}= -6H\omega_B^2$, which in turn follow from the linear auxiliary relation $({\rm D}_bB_a)^{\cdot}=-3H{\rm D}_bB_a$. The latter is obtained after combining the linear commutation law (A.2.2) of~\cite{BMT} with the linearised magnetic induction equation (i.e.~with $\dot{B}_a=-2HB_a$ -- see expression (\ref{Faraday}) in \S~\ref{ssMHDCLs}). Note that, in the absence of matter pressure, the Alfv\'en speed has become the wave velocity as well. Also note that the magneto-curvature effects reverse when the (rescaled) 3-curvature scalar ($\mathcal{K}$) changes from positive to negative and vice versa.

Our next step is to harmonically decompose Eq.~(\ref{D''}), by introducing the standard scalar harmonics functions $\mathcal{Q}^{(n)}$, with $\dot{\mathcal{Q}}^{(n)}=0$ and ${\rm D}^2\mathcal{Q}^{(n)}=-(n/a)^2\mathcal{Q}^{(n)}$. Then, setting $\Delta=\sum_n\Delta_{(n)}\mathcal{Q}^{(n)}$, $\mathcal{K}=\sum_n\mathcal{K}_{(n)}\mathcal{Q}^{(n)}$ and $(\sigma_B^2-\omega_B^2)= \sum_n(\sigma_B^2-\omega_B^2)_{(n)}\mathcal{Q}^{(n)}$, with ${\rm D}_a\Delta_{(n)}=0={\rm D}_a\mathcal{K}_{(n)}={\rm D}_a(\sigma_B^2-\omega_B^2)_{(n)}$, arrive at
\begin{eqnarray}\label{D harmonic}
\nonumber \ddot{\Delta}_{(n)}&=& -2H\dot{\Delta}_{(n)}+\frac{1}{2}\,\bar{\rho}\left[1-\frac{4}{9}\,c_{\rm a}^2 \left(\frac{\lambda_H}{\lambda_n}\right)^2\right]\Delta_{(n)}+ \frac{2}{3}\,\bar{\rho}c_{\rm a}^2\mathcal{K}_{(n)} \\&&
+4\left[1+\frac{1}{6}\left(\frac{\lambda_H}{\lambda_n}\right)^2\right] \left(\Sigma_B^2-\Omega_B^2\right)_{(n)}\,,
\end{eqnarray}
where $\lambda_H=1/H$ is the Hubble radius, $\lambda_n=a/n$ is the physical scale of the perturbation (with $n$ being the comoving wavenumber), while $\Sigma_B^2=a^2\sigma_B^2$ and $\Omega_B^2=a^2\omega_B^2$ define the rescaled magnetic shear and the magnetic voricity respectively. As expected, in the absence of these stresses, expressions (\ref{D''}) and (\ref{D harmonic}) reduce to the wavelike formula obtained in~\cite{BMT} (see Eq.~(7.4.5) there). Also, for a direct comparison between (\ref{D''}) and its Newtonian analogue, we refer the reader to Eq.~(27) in~\cite{VT}.

The second term on the right-hand side of (\ref{D harmonic}) conveys the opposing action of gravity and (magnetic) pressure. These effects cancel each other out (and the aforementioned term goes to zero) at a specific wavelength, which is given by
\begin{equation}\label{Jeans}
\lambda= \lambda_J= \frac{2}{3}\,c_{\rm a}\lambda_H
\end{equation}
and marks the (purely magnetic) Jeans length~\cite{TM,BMT}. On scales much larger than $\lambda_J$, gravity prevails and the perturbations grow. When $\lambda_n\ll\lambda_J$, however, the (positive) pressure of the $B$-field dominates and prevents the perturbations from growing (see \S~\ref{sSs} next).

\section{Linear solutions}\label{sSs}
We will examine the magnetic effects on the evolution of density perturbations in three steps. First, we will allow the magnetic pressure to act alone. Then, we will consider the simultaneous action of magnetic pressure and tension, leaving the role of the magneto-curvature coupling last.

\subsection{Magnetic pressure effects}
Without the magnetic tension terms, which include the magneto-curvature stresses as well, Eq.~(\ref{D harmonic}) reduces to
\begin{equation}\label{dif1.1}
\ddot{\Delta}_{(n)}=-2H\dot{\Delta}_{(n)}+\frac{1}{2}\,\bar{\rho} \left[1-\left(\frac{\lambda_J}{\lambda_n}\right)^2\right] \Delta_{(n)}\,,
\end{equation}
having substituted for the (magnetic) Jean's length from definition (\ref{Jeans}). During the dust era $a\propto t^{2/3}$, $H=2/3t$ and $\bar{\rho}=4/3t^2$. At the same time, the scale-ratio $\alpha=\lambda_J/\lambda_n$, which carries the magnetic effects, remains constant (recall that $c_{\rm a}\propto t^{-1/3}$ and $\lambda_n\propto t^{2/3}$ after equipartition, while $\lambda_H\propto t$ always and $\lambda_J=c_{\rm a}\lambda_H$). Then, the above differential equation assumes the form
\begin{equation}\label{dif1.2}
\frac{{\rm d}^2\Delta_{(n)}}{{\rm d}t^2}= -\frac{4}{3t}\,\frac{{\rm d}\Delta_{(n)}}{{\rm d}t}+ \frac{2}{3t^2}\left(1-\alpha^2\right)\Delta_{(n)}
\end{equation}
and accepts the power-law solution
\begin{equation}\label{sol1}
\Delta_{(n)}= C_1\,t^{s_1}+ C_2\,t^{s_2}\,,
\end{equation}
with $s_{1,2}=-[1\mp\sqrt{25-24\alpha^2}]/6$. Therefore, in the absence of the $B$-field (i.e.~when $\alpha=0$), we recover the standard non-magnetised solution for linear density perturbations in the dust era (i.e.~$s_{1,2}=-1, 2/3$ -- e.g.~see~\cite{TCM}). On the other hand, recalling that $\alpha^2=(4c_{\rm a}^2/9) (\lambda_H/\lambda_n)^2$, we deduce that the magnetic pressure inhibits the growth of these distortions by an amount proportional to the Alfv\'en-speed squared.\footnote{Analogous magnetic effects on the linear evolution of density perturbations were also observed during the radiation epoch, in solutions where only the pressure of the $B$-field was accounted for (see~\cite{TB2,BMT} for details).} Moreover, the impact of the aforementioned effect is scale-dependent. We will therefore consider the following three characteristic cases:

\begin{itemize}
  \item $\lambda_n\gg\lambda_J$: On scales much larger than the magnetic Jean's length, we have $\alpha= \lambda_J/\lambda_n\ll1$ and therefore solution (\ref{sol1}) reduces to
  \begin{equation}\label{1a}
  \Delta= C_1\,t^{2/3}+ C_2\,t^{-1}
  \end{equation}
  We have thus recovered the standard non-magnetised solution, which implies that the magnetic pressure has no effect on large scales.
  \item $\lambda_n\ll\lambda_J$: Here, $\alpha= \lambda_J/\lambda_n\gg1$, in which case solution (\ref{sol1}) takes the (imaginary) form
  \begin{equation}\label{1b}
  \Delta_{(n)}= t^{-1/6}\left(C_1\,t^{\imath\alpha\sqrt{2/3}}+ C_2\,t^{-\imath\alpha\sqrt{2/3}}\right)\,.
  \end{equation}
  Consequently, on small scales, the magnetic pressure dominates forcing the perturbations to oscillate (with amplitude that decreases as $t^{-1/6}$).
  \item $\lambda_n=\lambda_J$: At the $\alpha= \lambda_J/\lambda_n=1$ threshold the last term on the right-hand side of Eq.~(\ref{dif1.2}) vanishes and solution (\ref{sol1}) reads
  \begin{equation}\label{1c}
  \Delta= C_1+ C_2\,t^{-1/3}\,.
  \end{equation}
  In other words, on wavelengths equal to the Jeans length, the magnetic pressure balances the gravitational pull of the matter and the perturbations maintain constant amplitude.
\end{itemize}
Overall, the effects of the field's pressure are only felt on scales close and below the magnetic Jeans length. On larger wavelengths, the perturbations grow as if there was no $B$-field present. These results are identical to those obtained in the Newtonian study of~\cite{VT} and very close (both qualitatively and quantitatively) to those of the earlier relativistic treatments~\cite{TB1,BMT}.

\subsection{Combined pressure and tension effects}\label{ssCPTEs}
The opposite nature of the magnetic-pressure contribution and that of the field's tension (i.e.~positive vs negative) indicate that these two agents may act against each other. Here, we will attempt to clarify the matter by considering the combined effect of pressure and tension on the evolution of linear density perturbations. It should be noted, however, that the magneto-curvature stresses (which are also due to the magnetic tension) will remain switched off. Then, the density gradients are monitored by the linear system
\begin{equation}\label{D'' p+t}
\ddot{\Delta}_{(n)}= -2H\dot{\Delta}_{(n)}+ \frac{1}{2}\,\bar{\rho}\left[1 -\left(\frac{\lambda_J}{\lambda_n}\right)^2\right]\Delta_{(n)}+ 4\left[1+{1\over6}\left(\frac{\lambda_H}{\lambda_n}\right)^2\right] \left(\Sigma_B{}^2-\Omega_B{}^2\right)_{(n)}
\end{equation}
and
\begin{equation}\label{S}
\left(\Sigma_B{}^2-\Omega_B{}^2\right)^{\cdot}= -4H\left(\Sigma_B{}^2-\Omega_B{}^2\right)\,.
\end{equation}
The latter implies that $\Sigma_B$, $\Omega_B\propto a^{-2}$ on all scales and follows from the fact that $\Sigma_B=a\sigma_B$ and $\Omega_B=a\omega_B$, with $\sigma_B$, $\omega_B\propto a^{-3}$ (see Eqs.~(\ref{D''}) and (\ref{D harmonic}) in \S~\ref{ssW-LE}). Given that $H=2/3t$ and $\bar{\rho}=4/3t^2$ after equilibrium, the above recast as
\begin{equation}\label{d2Delta}
\frac{{\rm d}^2\Delta_{(n)}}{{\rm d}t^2}= -\frac{4}{3t}\frac{{\rm d}\Delta_{(n)}}{{\rm d}t}+ \frac{2}{3t^2}\left(1-\alpha^2\right)\Delta_{(n)}+ 4\left[1+{1\over6}\,\beta^2\left({t\over t_0}\right)^{2/3}\right] \left(\Sigma_B{}^2-\Omega_B{}^2\right)_{(n)}
\end{equation}
and
\begin{equation}\label{dSigma}
\frac{{\rm d}}{{\rm d}t}\left(\Sigma_B{}^2-\Omega_B{}^2\right)= -\frac{8}{3t}\left(\Sigma_B{}^2-\Omega_B{}^2\right)\,,
\end{equation}
respectively. As before, $\alpha=\lambda_J/\lambda_n=$~constant after equipartition, while $\beta=(\lambda_H/\lambda_n)_0=$~constant  determines the physical scale of the perturbations at the start of the dust era. When $\alpha\neq1,\sqrt{2/3}$, the system of (\ref{d2Delta}) and (\ref{dSigma}) solves analytically giving
\begin{equation}\label{gensol}
\Delta_{(n)}= C_1\,t^{s_1}+ C_2\,t^{s_2}+ C_3\left[\frac{\beta^2}{6(\alpha^2-1)} +\frac{1}{\alpha^2-{2/3}}\left({t_0\over t}\right)^{2/3}\right]\,,
\end{equation}
where $s_{1,2}=-[1\mp\sqrt{25-24\alpha^2}]/6$ exactly as before (see solution (\ref{sol1})). Consequently, the introduction of the tension stresses has added two extra modes (one constant and one decaying) to the linear evolution of magnetised density perturbations. Then, depending on the scale of the perturbation, we may consider the cases:
\begin{itemize}
  \item $\lambda_n\gg\lambda_J$: In this case $\alpha\ll1$ and the above solution reduces to\footnote{Although we use the same symbols for the integration constants in all our solutions, these generally differ.}
      \begin{equation}\label{s2.1}
       \Delta= C_1\,t^{2/3}+ C_2\,t^{-1}+ C_3+ C_4\,t^{-2/3}\,.
       \end{equation}
       Hence, on scales much larger than the magnetic Jeans length, the incorporation of the field's tension has not changed the standard picture. The density perturbations keep growing as $\Delta\propto t^{2/3}$, like their magnetic-free counterparts (compare to solution (\ref{1a})).
  \item $\lambda_n\ll\lambda_J$: Here $\alpha\gg1$ and expression (\ref{gensol}) becomes
      \begin{equation}\label{s2.2}
      \Delta_{(n)}=  t^{-1/6}\left(C_1\,t^{\imath\alpha\sqrt{2/3}}+ C_2\,t^{-\imath\alpha\sqrt{2/3}}\right)+ C_3+ C_4\,t^{-2/3}\,.
      \end{equation}
      As in solution (\ref{1b}) before, on small scales the magnetic pressure still forces the perturbations to oscillate with an amplitude that drops as $t^{-{1/6}}$. This time, however, the oscillations do not decay to zero but to a finite constant value that depends on the initial conditions.
  \item $\lambda_n=\lambda_J$: This special case corresponds to $\alpha=1$, when we can no longer use solution (\ref{gensol}). Instead, setting $\alpha=1$ into Eq.~(\ref{d2Delta}), the system of (\ref{d2Delta}), (\ref{dSigma}) gives
      \begin{equation}\label{s2.3}
      \Delta= C_1\ln t+ C_2+ C_3\,t^{-1/3}+ C_4\,t^{-2/3}\,.
      \end{equation}
      Therefore, at the magnetic Jeans length, where the field's pressure cancels out the gravitational pull of the matter, the magnetic tension becomes the sole player, takes over and leads to a weak (logarithmic) growth of the perturbations. Recall that, in the absence of tension stresses, perturbations with wavelength equal to the Jeans length remain constant (see solution (\ref{1c}) before). The growth seen in solution (\ref{s2.3}) demonstrates the opposing action between the field's pressure and tension on the linear evolution of density perturbations, which lies at the core of this investigation.
  \item $\lambda_n=\sqrt{3/2}\,\lambda_J$: This is our second special case, corresponding to $\alpha=\sqrt{2/3}$ and to $\beta^2(t/t_0)^{2/3}\gg1$, in which case the system (\ref{d2Delta}) and (\ref{dSigma}) accepts the solution
      \begin{equation}\label{s2.4}
      \Delta= C_1\,t^{1/3}+ C_2+ C_3\,t^{-2/3}\,.
      \end{equation}
       Consequently on scales that are only slightly larger than the magnetic Jean's length the perturbations grow as $t^{1/3}$, instead of following the $\Delta\propto t^{2/3}$-law associated with much larger wavelengths (see solutions (\ref{1a}) and (\ref{s2.1})). This implies that the growth-rate of density perturbations increases gradually as we move on to scales progressively larger than $\lambda_J$ and the overall magnetic effect weakens (which is to be expected).\footnote{A closer look into the study of~\cite{VT} reveals that solutions (\ref{s2.3}) and (\ref{s2.4}) reside in the Newtonian equations as well, although not as distinct special cases, which is probably the reason they were not identified there.}
\end{itemize}

\subsection{Including the magneto-curvature effects}\label{ssIM-CEs}
In order to incorporate the magneto-curvature stresses seen in Eqs.~(\ref{D''}) and (\ref{D harmonic}) into our solutions, we need to involve the evolution formula of the rescaled 3-Ricci scalar (see expression (\ref{lcK})). Then, taking the time derivative of (\ref{D''}) and using (\ref{lcK}) and (\ref{S}), we arrive at the differential equation
\begin{equation}\label{D'''}
\dddot{\Delta}= - 6H\ddot{\Delta}- {7\over6}\,\bar{\rho}\dot{\Delta}+ {1\over2}\,H\bar{\rho}\Delta+ {2\over3}\,Hc_{\rm a}^2{\rm D}^2\Delta+ {2\over3}\,c_{\rm a}^2{\rm D}^2\dot{\Delta}- {2H\over\bar{\rho}}\,{\rm D}^2 \left(\Sigma_B^2-\Omega_B^2\right)\,.
\end{equation}
Harmonically decomposed the above reads
\begin{eqnarray}\label{D''' hd}
\nonumber \dddot{\Delta}_{(n)}&=& -6H\ddot{\Delta}_{(n)}- {7\over6}\,\bar{\rho} \left(1+\frac{3}{7}\,\alpha^2\right)\dot{\Delta}_{(n)}+ \frac{1}{2}\,H\bar{\rho}\left(1-\alpha^2\right)\Delta_{(n)} \nonumber\\ &&+{2\over3}\,H\beta^2\left({t\over t_0}\right)^{2/3} \left(\Sigma_B^2-\Omega_B^2\right)_{(n)}\,,
\end{eqnarray}
with $\alpha=\lambda_J/\lambda_n=$~constant and $\beta=(\lambda_H/\lambda_n)_0$. After equipartition, the above differential equation takes the form
\begin{eqnarray}\label{D''' eq}
\nonumber \frac{{\rm d}^3\Delta_{(n)}}{{\rm d}t^3}&=& -\frac{4}{t}\,\frac{{\rm d}^2\Delta_{(n)}}{{\rm d}t^2}- \frac{14}{9t^2}\,\left(1+\frac{3}{7}\,\alpha^2\right)\frac{{\rm d}\Delta_{(n)}}{{\rm d}t}+ \frac{4}{9t^3}\left(1-\alpha^2\right)\Delta_{(n)} \\&& +\frac{4}{9t}\,\beta^2\left({t\over t_0}\right)^{2/3} \left(\Sigma_B^2-\Omega_B^2\right)_{(n)}\,.
\end{eqnarray}
Finally, when $\alpha\neq1$, the system of (\ref{dSigma}) and (\ref{D''' eq}) solves to give
\begin{equation}\label{gen sol3}
\Delta_{(n)}= C_1\,t^{s_1}+ C_2\,t^{s_2}+ {1\over\alpha^2-1}\left(C_3+C_4\,t^{-2/3}\right)\,.
\end{equation}
with $s_{1,2}=-[1\mp\sqrt{25-24\alpha^2}]/6$. When $\alpha=1$, on the other hand we obtain
\begin{equation}
\Delta= C_1\,\ln t+ C_2- C_3\,t^{-1/3}+ C_4\,t^{-2/3}\,.
\end{equation}
For all practical purposes, the above results are identical to solutions (\ref{gensol}) and (\ref{s2.3}), implying that the inclusion of the magneto-curvature effects does not alter the linear evolution of the density perturbations. This is not surprising, since the spatial flatness of the FRW background ensures that the magneto-curvature stresses are too weak to make a noticeable difference.

\section{Discussion}\label{sD}
With the exception of the Cosmic Microwave Background (CMB), magnetic fields have been observed nearly everywhere in the cosmos. The idea of primordial magnetism has also been gaining ground because it could in principle explain all the large-scale $B$-fields seen in the universe today. If present, cosmological magnetic fields could have played a role during structure formation, since they can in principle generate and affect the evolution of all types of perturbations, namely scalar, vector and tensor distortions (see \S~\ref{ssTIs}). When it comes to scalar (density) perturbations, however, half of the magnetic effects are excluded, since all the available cosmological studies (with the exception of~\cite{VT} -- to the best of our knowledge) account only for field's pressure and bypass the magnetic tension.\footnote{In astrophysics the implications of the magnetic tension have been investigated in a number of studies looking at the physics of star formation, accretion discs and compact stars (e.g.~see~\cite{NCVR}-\cite{MM} and references therein).} Moreover, technically speaking, it is more straightforward to obtain analytic solutions before rather than after equipartition. The main difficulty comes from the Alfv\'en speed, which is constant throughout the radiation era but acquires a time-dependence after equilibrium. As result, the available dust-epoch solutions were obtained after imposing certain simplifying assumptions~\cite{TB1}-\cite{BMT}. In the present work we re-examine the magnetic implications for the evolution of baryonic density perturbations and try to address both of the aforementioned issues. Our study uses full general relativity, incorporates the effects of the field's tension and focuses on the post-recombination universe. The aim was to refine and extend previous relativistic studies, as well as provide a direct comparison with the existing Newtonian treatments of the issue. Above all, however, we wanted to investigate and reveal the as yet unknown role of the magnetic tension.

At the centre of our analysis is the wave-like equation monitoring the linear evolution of magnetised density perturbations. In contrast to previous approaches, this formula carries the effects of the magnetic tension, in addition to those of the field's pressure. After equipartition, the latter is the sole source of support against the gravitational pull of the matter. This leads to a purely magnetic Jeans length, which means that the magnetic pressure could in principle determine the first gravitationally bound formations. The tension stresses, on the other hand, are triggered by the elasticity of the field lines and by their natural tendency to react against any agent that distorts them from equilibrium. Among these are the magneto-curvature stresses, which result from the purely geometrical coupling between the $B$-field and the spatial geometry of the host spacetime. We have incorporated all the aforementioned effects into our analytic solutions in three successive steps of increasing inclusiveness. At first, we only considered the effects of the field's pressure, in which case our results were in full agreement with those of the previous Newtonian study. We then also accounted for the role of the magnetic tension and finally, to complete the picture, we incorporated the magneto-curvature stresses as well. Our results showed that the field's pressure and tension act against each other. The magnetic pressure, in particular, inhibits the growth of the perturbations, while the tension tends to enhance it. These effects were also found to be scale-dependent, with the pressure dominating well inside the (purely magnetic) Jeans length and with the tension taking over near the Jeans threshold. On much larger wavelengths, on the other hand, neither of these agents had a measurable effect and the perturbations evolved unaffected by the field's presence

More specifically, well inside the magnetic Jeans length and in the absence of any tension input, we found that the field's pressure forces the perturbations to oscillate with an amplitude that decreases as $\Delta\propto t^{-1/6}$ and decays (asymptotically) to zero. When the magnetic tension was included the oscillations still decayed (at the same rate), though now to a finite value instead of zero. Near the Jeans length the support of the field's pressure and the gravitational pull of the matter cancel each other out, thus leaving the magnetic tension as the sole player. This resulted into a slow logarithmic growth for the density perturbations, which revealed the (as yet unknown) opposing action of the aforementioned two magnetic agents on the linear evolution of density gradients. We expect an analogous effect near the Jeans length during the radiation era as well. Qualitatively speaking, the role of the magnetic tension demonstrated how versatile and unconventional the $B$-fields can be. Quantitatively, the tension effects were relatively weak because their contribution decays quickly (faster than that of the field's pressure) with the universal expansion. Nevertheless, it is conceivable that there can be physical situations where the field's tension could play a more prominent role. This should probably happen in the nonlinear phase of structure formation on scales considerably smaller than the magnetic Jeans length and more likely during the (typically) anisotropic collapse of a magnetised protogalactic cloud. Given the complexity of the nonlinear regime, however, one should have to employ numerical methods to complement the analytical work. Beyond the Jeans length, the overall magnetic effect was found to gradually fade away and the standard (non-magnetised) linear growth-rate of density perturbations was eventually re-established. Finally, the magneto-curvature stresses (which also result from the field's tension) were found to be too weak to leave a measurable imprint. This was largely expected, however, given the (assumed) spatial flatness of our FRW background.\footnote{In order to study the coupling between magnetism and spacetime geometry in detail and to investigate its potential implications in depth, one needs to allow for FRW backgrounds with nonzero spatial curvature.} What is particularly interesting about these stresses, is that their effect reverses depending on the sign of the spatial curvature (i.e.~on whether it is positive or negative -- see Eq.~(\ref{D''}) in \S~\ref{ssW-LE}). Therefore, if these magneto-geometrical effects were to be detected, they should also provide information about the universe's spatial geometry.\\

\textbf{Acknowledgments} JDB was supported by the the Science and Technology Facilities Council (STFC) of the United Kingdom.

\end{document}